\documentclass[11pt]{article}
\usepackage{moriond,epsfig}
\begin{document}
\vspace*{4cm}
\title{ON THE PROSPECTS OF TAU NEUTRINO ASTRONOMY IN GEV ENERGIES AND BEYOND}

\author{H. ATHAR$^{1,2}$, FEI-FAN LEE$^2$ and GUEY-LIN LIN$^2$ }
\address{$^1$Physics Division, National Center for Theoretical Sciences,
Hsinchu 300, Taiwan\\
$^2$Institute of Physics, National Chiao-Tung University, Hsinchu
300, Taiwan}

\maketitle
\abstracts{We point out the opportunity of tau neutrino
astronomy for neutrino energies of the order of 10 GeV to 10$^{3}$ GeV.
In this energy range, it is demonstrated that the flavor dependence
in the background atmospheric neutrino flux leads to drastically
different prospects between the observation of astrophysical muon
neutrinos and that of astrophysical tau neutrinos. Taking the
galactic-plane neutrino flux as a targeted astrophysical source,
we found that the galactic-plane tau neutrino flux dominates over
the atmospheric tau neutrino flux for neutrino energies beyond 10
GeV. Hence the galactic-plane can in principle be seen through tau
neutrinos with energies greater than 10 GeV. In a sharp contrast,
the galactic-plane muon neutrino flux is overwhelmed by its
atmospheric background until the energy of $10^{6}$ GeV.}
\section{Introduction}
The atmospheric $\nu_{\mu}\to \nu_{\tau}$ oscillations established by
the Super-Kamiokande (SK) detector ensures that a non-negligible
$\nu_{\tau}$ flux reaches the Earth. The updated SK analysis of the
atmospheric neutrino data gives \cite{Ashie:2004mr}
\begin{equation}
 1.9\cdot 10^{-3}\,\,\, {\rm eV}^{2}< \delta m^{2} < 3.0\cdot
 10^{-3}\,\,\, {\rm eV}^{2},  \ \sin^{2}2\theta >0.9. \label{range23}
\end{equation}
This is a  $90\% \, {\rm C.L.}$ range with the best fit values given
by $\sin^{2}2\theta =1$ and $\delta m^{2}=2.4\cdot 10^{-3}\, \, \,
{\rm eV}^{2}$ respectively. The tau neutrinos arising from the above
$\nu_{\mu}\to \nu_{\tau}$ oscillations are presently identified on
the statistical basis \cite{icrr2003}. On the other 
hand  \cite{Kajita:2000mr}, the total
number of observed non-tau neutrinos from various detectors are
already greater than $10^{4}$ with energies ranging from $10^{-1}$
GeV to $10^{3}$ GeV. It is essential to develop
efficient techniques for the identification of tau neutrinos.

For $E\geq  10^{6}$ GeV, the $\nu_{\tau}$ can be detected by
the {\em double bang} showers of the tau lepton, which is produced by the
neutrino nucleon charge current (CC) scattering
\cite{Learned:1994wg,Athar:2000rx}. In this case, the two showers are
separated by the tau lepton decay length, which is roughly $50$ m at
$E_{\tau}=10^6$ GeV. For $E < 10^{6}$ GeV, 
 one relies on the distinctive properties of the tau neutrino
induced showers to detect $\nu_{\tau}$. 
 At low energy, the showers produced by 
 CC $\nu_{\tau}$ interaction and the subsequent tau decay
practically coincide. Treating these two showers as one, it is found
that a much higher fraction of tau neutrino energy deposits in the
form of showers than either in the $\nu_{\mu}$ CC interactions or
in the neutral current (NC) interactions of any neutrino flavor.
Hence, the footprints of $\nu_{\tau}$ might be identified by the study of 
energy spectrum of shower events \cite{Stanev:1999ki}. 
 The ratio of the combined electromagnetic and
hadronic shower event rate to the muon event rate is also a sensitive
probe to tau neutrino appearance due to oscillations
\cite{Dutta:2000jv}. The identification of
$\nu_{\tau}$ appearance using shower properties requires that the
flux of $\nu_{\tau}$ is comparable to those of other neutrino
flavors. It is a challenging task to identify
$\nu_{\tau}$ if other neutrino flavors  dominate in flux.
Nevertheless, the tau neutrino astronomy will be rather promising if the
above tasks can be carried out, as we shall demonstrate in the
following.
\section{The Prospects of Tau Neutrino Astronomy}
We illustrate the opportunity of tau neutrino astronomy by the
possible detection of astrophysical $\nu_{\tau}$ from the galactic
plane direction. To do this, we estimate the flux of galactic $\nu_{\tau}$ and
calculate the background atmospheric $\nu_{\tau}$ flux in detail.
The tau neutrino flux arriving at the detector on Earth, after
traversing a distance $L$ can be written as
\begin{equation}
\label{osc}
 \phi_{\nu_{\tau}}^{\rm tot}(E)=P(E)\cdot
 \phi_{\nu_{\mu}}(E)+(1-P(E))\cdot \phi_{\nu_{\tau}}(E),
\end{equation}
where $\phi_{\nu_{\mu}}(E)$ and $\phi_{\nu_{\tau}}(E)$ are intrinsic
muon neutrino and tau neutrino fluxes respectively, $P(E)\equiv
P(\nu_{\mu}\to \nu_{\tau})=\sin^2 2\theta\cdot \sin^2(L/L_{\rm
osc})$ is the $\nu_{\mu}\to \nu_{\tau}$ oscillation probability with
the oscillation length given by $L_{\rm osc}=4E/\delta m^2$.
\subsection{The galactic-plane tau neutrino flux}
One calculates the intrinsic galactic-plane $\nu_{\mu}$ and
$\nu_{\tau}$ fluxes by considering the collisions of incident
cosmic-ray protons with the interstellar medium. The fluxes are given
by
\begin{equation}
\label{gala}
 \phi_{\nu}(E)= Rn_p\int_{E}^{\infty}
 \mbox{d}E_p \; \phi_{p}(E_{p})\;
 \frac{\mbox{d}\sigma_{pp \to \nu +Y}}{\mbox{d}E},
\end{equation}
where $E_p$ is the energy of incident cosmic-ray proton,
$\mbox{d}\sigma_{pp \to \nu +Y}/\mbox{d}E$ is the neutrino energy
spectrum in the $pp$ collisions, $R$ the typical distance in the
galaxy along the galactic plane, which we take as $10$ kpc (1 pc
$\sim 3\cdot 10^{16}$ m). The density of the interstellar medium
$n_p$ along the galactic plane is taken to be $\sim$ 1 proton per
cm$^3$. The primary cosmic-ray proton flux, $\phi_p(E_p)\equiv {\rm
d}N_p/{\rm d}E_p$ for $E_{p}\leq 10^{4}$ GeV, is given by \cite{Gaisser:2002jj}
\begin{equation}
\label{cosmic}
 \phi_p(E_{p})=1.49\cdot \left(E_{p}+2.15\cdot
 \exp(-0.21\sqrt{E_{p}})\right)^{-2.74},
\end{equation}
in units of cm$^{-2}$s$^{-1}$sr$^{-1}$GeV$^{-1}$.

The flux of galactic-plane muon neutrino arises from the two-body
$\pi$ decays and the subsequent three-body muon decays. The
differential cross section for $p+p\to \pi+X$ is model dependent. We
adopt the parameterization in \cite{Gaisser:2001sd} for such a cross
section.

The galactic-plane $\nu_{\tau}$ flux arises from the productions and
decays of the $D_s$ mesons. It has been found to be rather
suppressed compared to the corresponding $\nu_{\mu}$ flux
\cite{Athar:2001jw}. Clearly the total galactic-plane tau neutrino
flux, $\phi_{\nu_{\tau}}^{\rm tot}(E)$, is dominated by the
$\nu_{\mu}\to \nu_{\tau}$ oscillations indicated by the term
$P(E)\cdot \phi_{\nu_{\mu}}(E)$ in Eq.~(\ref{osc}). We have
\cite{Athar:2004um}
\begin{equation}
\phi_{\nu_{\tau}}^{\rm tot}(E)=2\cdot 10^{-5}\left(\frac{E}{\rm
GeV}\right)^{-1.64},
\end{equation}
in units of cm$^{-2}$s$^{-1}$sr$^{-1}$ for $1 \ {\rm GeV}\leq E\leq
10^3 \ {\rm GeV}$.
\subsection{The atmospheric tau neutrino flux}
We follow a semi-analytic approach \cite{Gaisser:2001sd} for
computing the flux of intrinsic atmospheric $\nu_{\mu}$ which could
oscillate into $\nu_{\tau}$. For $\pi$-decay contribution, the flux
in the notation of  \cite{Gaisser:2001sd} reads:
\begin{eqnarray}
\label{atm-nu}
 \frac{\mbox{d}^2N^{\pi}_{\nu_{\mu}}(E,\xi,X)}{\mbox{d}E\mbox{d}X}&=&\int_{E}^{\infty}
 \mbox{d}E_N\int_{E}^{E_{N}}
 \mbox{d}E_{\pi}\frac{\Theta(E_{\pi}-\frac{E}{1-\gamma_{\pi}})}{d_{\pi}E_{\pi}(1-\gamma_{\pi})}
 \int_0^X
 \frac{\mbox{d}X'}{\lambda_N}P_{\pi}(E_{\pi},X,X')\nonumber \\
 & &\times \frac{1}{E_{\pi}}F_{N\pi}(E_{\pi},E_N)
 \exp \left(-\frac{X'}{\Lambda_N}\right)\phi_N(E_N).
\end{eqnarray}
We only consider the proton component of $\phi_N$, which is given by
Eq.~(\ref{cosmic}). 
The kaon contribution to the atmospheric $\nu_{\mu}$ flux can be
computed in the similar way. We note that charmed-hadron decays also
contribute to the atmospheric $\nu_{\mu}$ flux. However, these
contributions are negligible for $E < 10^5$ GeV.

The intrinsic atmospheric $\nu_{\tau}$ flux is reliably calculable using 
 the perturbative QCD \cite{Pasquali:1998xf}. One writes the flux as
\begin{equation}
\label{atm-tau}
 \frac{\mbox{d}^2N_{\nu_{\mu}}(E,X)}{\mbox{d}E\mbox{d}X}=
 \frac{Z_{pD_s}Z_{D_s\nu_{\tau}}}{1-Z_{pp}(E)}\cdot
 \frac{\exp(-X/\Lambda_p)\phi_p(E)}{\Lambda_p},
 \end{equation}
where the $Z$ moments on the RHS of the equation are defined by
\begin{equation}
\label{z-moment}
 Z_{ij}(E_j)\equiv \int_{E_j}^{\infty}{\mbox
 d}E_i\frac{\phi_i(E_i)}{\phi_i(E_j)}\frac{\lambda_i(E_j)}{\lambda_i(E_i)}
 \frac{{\mbox d}n_{iA\to jY}(E_i,E_j)}{{\mbox d}E_j},
\end{equation}
with ${\mbox d}n_{iA\to jY}(E_i,E_j)\equiv {\mbox d}\sigma_{iA\to
jY}(E_i,E_j)/\sigma_{iA}(E_i)$.
We have calculated the total atmospheric $\nu_{\tau}$ flux by
applying Eq.~(\ref{osc}) with $\phi_{\nu_{\mu,\tau}}(E)$ given by
$\mbox{d}^2N_{\nu_{\mu,\tau}}(E,X)/\mbox{d}E\mbox{d}X$ and
integrating over the slant depth $X$. For $\xi < 70^{\circ}$, the
oscillation probability $P(\nu_{\mu}\to \nu_{\tau})$ is calculated
using the relation $X=X_0\exp(-L\cos\xi/h_0)/\cos\xi$, with
$X_0=1030$ g/cm$^2$, $h_0=6.4$ km, and $L$ the linear distance from
the neutrino production point to the detector on Earth
\cite{Gaisser:1997eu}.

\begin{figure}[tbp]
\begin{center}
\psfig{figure=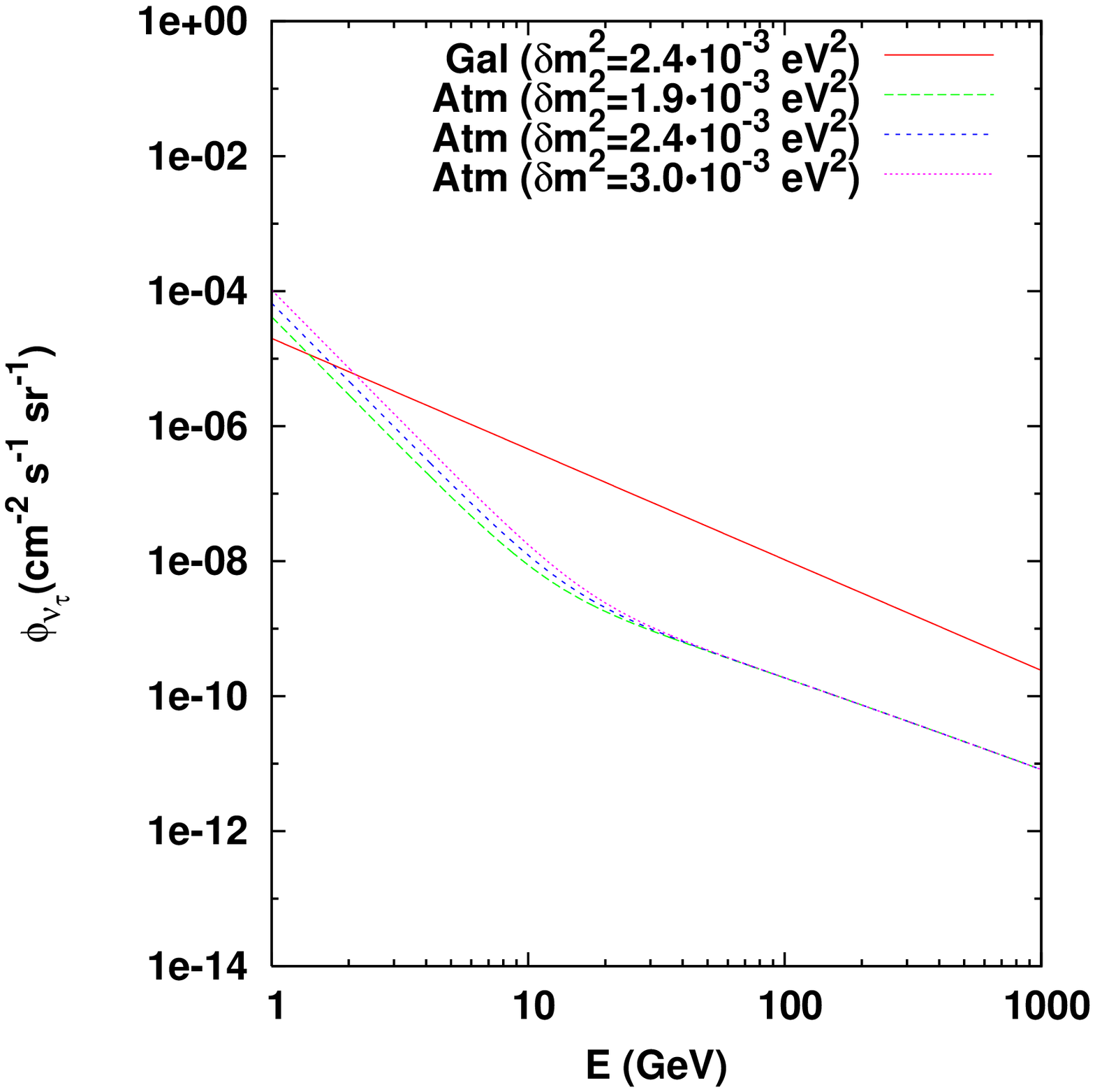,height=3.0in}
\psfig{figure=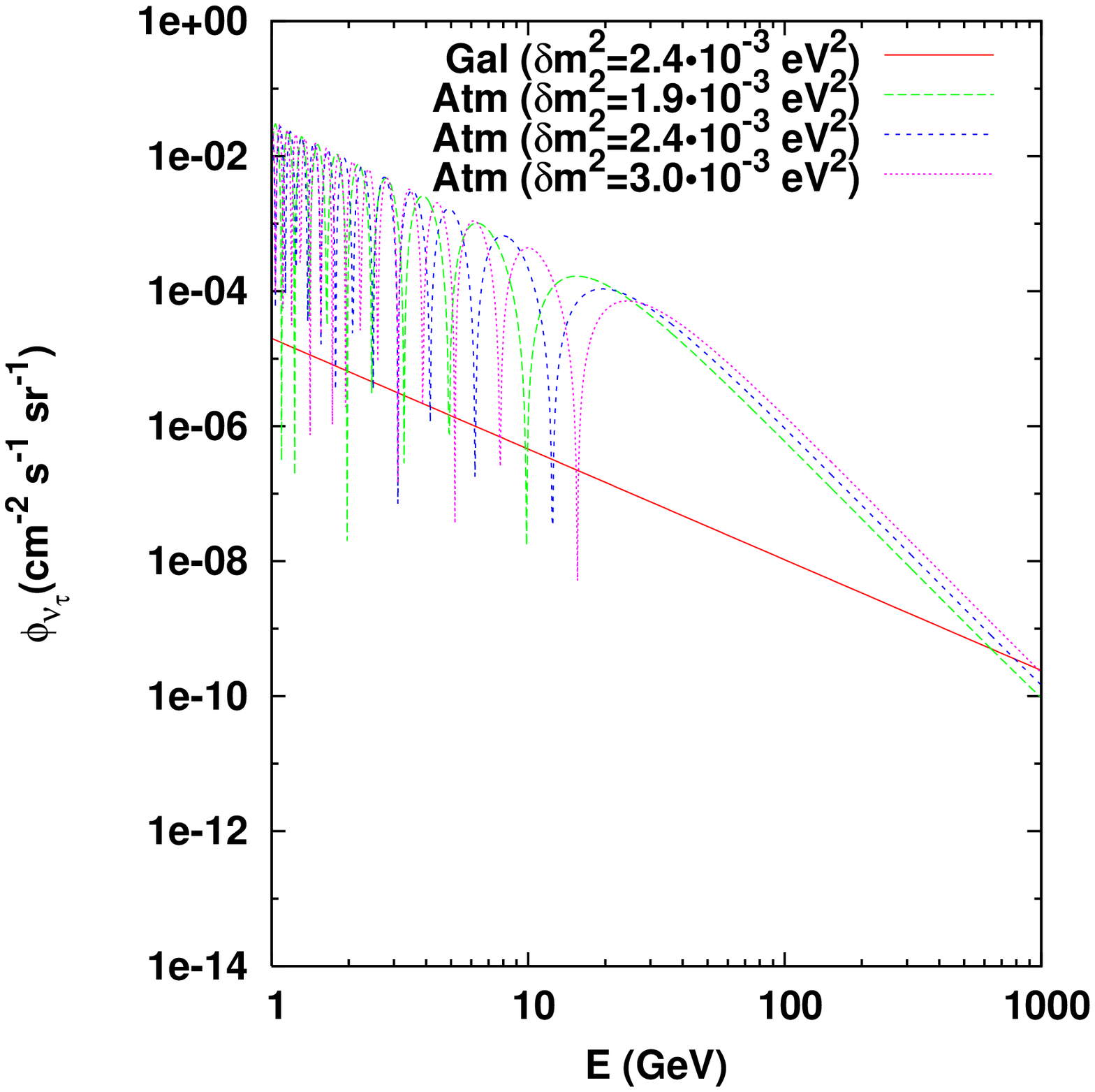,height=3.0in}
 \caption{The comparison of  
galactic-plane and the atmospheric $\nu_{\tau}$ fluxes in the presence
of  neutrino oscillations. The atmospheric $\nu_{\tau}$ fluxes in the
left figure are the downward, while those in the right figure are
 the upward. \label{fig:comparison}}
\end{center}
\end{figure}

The comparison of the galactic-plane and the atmospheric
$\nu_{\tau}$ fluxes is given in Fig.~\ref{fig:comparison}. The plot
on the left hand side compares the galactic-plane $\nu_{\tau }$ flux
and the downward atmospheric background flux. For $\delta
m^2=2.4\cdot 10^{-3}$ eV$^2$, $\sin^2 2\theta=1$, we find that both
fluxes {\tt cross} at $E=2.3$ GeV. It is seen that the atmospheric
$\nu_{\tau}$ flux is sensitive to the value of $\delta m^2$ for
$E\leq 20$ GeV. This flux also changes its slope at $E\approx 20$
GeV. Below $20$ GeV, the atmospheric $\nu_{\tau}$ flux predominantly
comes from the $\nu_{\mu}$ oscillations, i.e., $\phi^{\rm
tot}_{\nu_{\tau}}(E)\approx \phi_{\nu_{\mu}}(E)\cdot \sin^2
2\theta\cdot \sin^2(L/L_{\rm osc})$ following Eq.~(\ref{osc}). Since
$L_{\rm osc}\equiv 4E/\delta m^2\approx 330$ km for $E=1$ GeV and
$\delta m^2=2.4\cdot 10^{-3}$ eV$^2$, we approximate
$\sin^2(L/L_{\rm osc})$ with $(L/L_{\rm osc})^2$, so that $\phi^{\rm
tot}_{\nu_{\tau}}(E)\sim \phi_{\nu_{\mu}}(E)E^{-2}$. Because the
neutrino oscillation effect steepens the $\phi_{\nu_{\tau}}$
spectrum for $E \leq 20$ GeV, the slope change of
$\phi_{\nu_{\tau}}$ at $E\approx 20$ GeV is therefore significant.
The plot on the right hand side of Fig.~\ref{fig:comparison}
compares the galactic-plane $\nu_{\tau}$ flux and upward atmospheric
background flux. The two fluxes cross at $E=8\cdot 10^{2}$ GeV for the
best-fit neutrino oscillation parameters. We have also compared the
galactic-plane and atmospheric $\nu_{\tau}$ fluxes for several other
zenith angles. For instance, for the zenith angle $\xi=60^{\circ}$, these two
fluxes cross at $E=6.0$ GeV for $\delta m^2=2.4\cdot 10^{-3}$ eV$^2$
with the maximal mixing.

The above comparison of galactic-plane and atmospheric $\nu_{\tau}$
fluxes indicates a {\tt window of opportunity} for the tau neutrino
astronomy. Clearly, if $\nu_{\tau}$ can be identified from the
overwhelming $\nu_{\mu}$ background (in this case the atmospheric
$\nu_{\mu}$ flux), the galactic-plane can be seen through GeV energy
tau neutrinos in the downward directions. In the upward directions,
galactic-plane tau neutrinos are observable for $E\geq  10^{3}$ GeV.

It is instructive to also compare the galactic-plane and the atmospheric
$\nu_{\mu}$ fluxes. While the galactic-plane neutrino flux is flavor
independent, the atmospheric $\nu_{\mu}$ flux dominates its
$\nu_{\tau}$ counterpart. As a consequence, the crossing point of
galactic-plane and atmospheric $\nu_{\mu}$ fluxes is pushed up to
$5\cdot 10^5$ GeV, which is {\em  drastically different} from $\nu_{\tau}$ 
 case \cite{Athar:2004um,Athar:2003nc}. This is a general
situation in the $\nu_{\mu }$ astronomy where the opportunity for
observing  astrophysical neutrinos begins typically at $10^{6}$ GeV. In
contrast, with the future development of $\nu_{\tau}$
identification techniques,  the energy
threshold for the neutrino astronomy might significantly be lowered down.
\section*{Acknowledgments}
We thank the organizer for the invitation to present this work. H.A.
thanks Physics Division of NCTS for support. F.F.L. and G.L.L. are
supported by the National Science Council of Taiwan under the grant
number NSC 93-2112-M-009-001.
\section*{References}
\end{document}